\documentclass[twoside,a4paper]{article}
\usepackage{booktabs}
\usepackage{xspace,xcolor,svgcolor}
\usepackage{embedfile}
\usepackage{amssymb}
\usepackage{fullpage}
\usepackage{amsmath}
\usepackage{amsthm}
\usepackage{multirow}
\usepackage{rotating}
\usepackage{ifpdf}
\ifpdf%
\usepackage{ae} 
\usepackage{aeguill}
\fi
\usepackage[pdftex,backref=section]{hyperref}
\hypersetup{%
colorlinks,
citecolor=blue,
filecolor=black,
linkcolor=darkred,
urlcolor=teal
}
\usepackage[all]{hypcap} 

\usepackage{listings}
\lstdefinestyle{slp}{
 basicstyle=\ttfamily\footnotesize,
 breaklines=true,
 breakatwhitespace=true,
 breakindent=0pt,
 string=[d]{;},
 keywordstyle=\color{blue},
 showstringspaces=false,
 xleftmargin=2pt, xrightmargin=2pt,
 numbers=none
}
\usepackage[nameinlink,capitalize,noabbrev]{cleveref}
\newcommand{\K}{\mathbb{K}}

\newcommand{\bigO}[1]{\ensuremath{{\mathcal O}\!\left(#1\right)}}
\newcommand{\SLP}{{\textsc{slp}}\xspace}
\newcommand{\SLPs}{{\textsc{slp}s}\xspace}
\newcommand{\FMMA}[4]{\ensuremath{{\langle{{#1}{\times}{#2}{\times}{#3}{:}{#4}}\rangle}}}

\newcommand{\vectorization}[1]{\ensuremath{\textup{Vect}{(#1)}}}%
\newcommand{\matrixsize}[2]{\ensuremath{{{#1}{\times}{#2}}}}
\newcommand{\mat}[1]{\ensuremath{\mathsf{#1}}}%

\newenvironment{smatrix}{\left[\begin{smallmatrix}}{\end{smallmatrix}\right]}
\newcommand{\firstdim}{\ensuremath{m}}
\newcommand{\seconddim}{\ensuremath{k}}

\newcommand{\Transpose}[1]{\ensuremath{{{#1}}^{\intercal}}}

\theoremstyle{plain}

\newtheorem{lemma}{Lemma}[section]
\theoremstyle{definition}

\theoremstyle{remark}
\newtheorem{remark}{Remark}[section]

\newcommand{\plinoptgit}{https://github.com/jgdumas/plinopt}
\newcommand{\plinopt}{\href{\plinoptgit}{\textsc{PLinOpt}}\xspace}
\newcommand{\plinoptdata}[1]{\href{https://github.com/jgdumas/plinopt/tree/main/data}{\detokenize{#1}}}
\title{Fast matrix multiplication via recursive \FMMA{4}{4}{4}{48} algorithms into practice}

\newcommand{\email}[1]{\href{Milton:#1}{#1}}

\author{Jean-Guillaume Dumas, 
\email{Jean-Guillaume.Dumas@univ-grenoble-alpes.fr}%
\footnote{
Univ.\ Grenoble Alpes, Laboratoire Jean Kuntzmann, \textsc{umr} 5224 \textsc{cnrs},
Grenoble,
France
}
\and
Cl\'ement Pernet, 
\email{Clement.Pernet@univ-grenoble-alpes.fr}%
\footnote{
Grenoble \textsc{inp} -- \textsc{uga}, Laboratoire Jean Kuntzmann, \textsc{umr} 5224 \textsc{cnrs},
Grenoble,
France
}
\and
Alexandre Sedoglavic, 
\email{Alexandre.Sedoglavic@univ-lille.fr}
\footnote{
Universit\'e de Lille, Centrale Lille, \textsc{umr} \textsc{cnrs} 9189 \textsc{cris}t\textsc{al},
Lille,
France
}\\[-13pt]
\and
Joshua Stapleton, 
\email{Joshua.Stapleton.ai@gmail.com} 
\and
Petr Tichavsk{\'y}, 
\email{Petr.Tichavsky@utia.cas.cz}
\footnote{
Academy of Sciences of the Czech Republic, Institute of Information Theory and Automation, 
Praha,
Czech Republic%
}
}
\begin{document}
\maketitle
\begin{abstract}
We present a fast algorithm for multiplying two~\(\matrixsize{4}{4}\)
matrices using~\(48\) multiplications and~\(204\) other operations
(addition, subtraction or scaling by a constant)
over any ring containing an inverse of~\(2\).
Applied recursively, this algorithm reaches a cost bound with leading
term~\({\left(7+\frac{3}{8}\right)n^{\log_{4}\!{48}}}\).
Alternative basis decompositions of this algorithm further
reduce this to~\({\left(6+\frac{13}{32}\right)n^{\log_4\!{48}}}\),
with square change of bases, and then
to~\({\left(5+\frac{2}{3}\right)n^{\log_4\!{48}}}\),
using rectangular encodings with inner dimensions~\(24\) and~\(32\).
The conversion costs of these alternative basis variants
are included as lower-order terms.
\end{abstract}
\setcounter{tocdepth}{3}
\tableofcontents
\section{Introduction}
This paper focuses on reducing the leading constant factor in the complexity
bound for matrix multiplication, building on recent advances in
algorithmic design.
\par
Recently indeed, the number of scalar multiplications required to
multiply two~\(\matrixsize{4}{4}\) matrices was reduced
in~\cite{Fawzi:2022aa} from~\(49\)~(using two recursion levels of
Strassen's algorithm~\cite{strassen:1969}) to~\(47\) in
characteristic~\(2\).
More recently, it was reduced to~\(48\) in~\cite{alphaevolve} over the complex numbers, and then to~\(48\) over any ring containing an inverse of~\(2\) in~\cite{DPS25:444}.
Applied recursively, a~\(48\) multiplication algorithm has a cost bound with a leading exponent of~\({\log_{4}\!{48}=2+\log_{4}\!{3} \approx 2.7925}\).
\par
In this paper, we focus on the \emph{constant} of this leading term in the complexity bound.
The first straight-line program presented in~\cite{DPS25:444} requires~\(315\) scalar
additions, subtractions and scalings by a constant, resulting in a recursive algorithm whose cost's leading term is:
\begin{equation}
 {\left(1+\frac{315}{48-16}\right)n^{2+\log_{4}\!{3}}=\frac{347}{32}n^{2+\log_{4}\!{3}}=\left(10+\frac{27}{32}\right)n^{2+\log_{4}\!{3}}\approx{10.84375~n^{2.7925}}}.
\end{equation}
An alternative basis version of this algorithm, following the technique
of~\cite{Karstadt:2017aa,Beniamini:2019aa}, then has a cost with leading term~\({7n^{2+\log_{4}\!{3}}}\).
A variant of that algorithm with improved accuracy was presented
in~\cite{Dumas:2026ac} with a leading term
of~\({\left(9+\frac{7}{8}\right)n^{2+\log_{4}\!{3}}}\) (the best
currently known such variant leading term
is~\({\left(9+\frac{3}{16}\right)n^{2+\log_{4}\!{3}}}\),
or~\({\left(5+\frac{3}{4}\right)n^{2+\log_{4}\!{3}}}\) for an
alternative basis version).
\par
As a comparison with a different number of bilinear products,
Strassen-Winograd algorithm with~\(7\)~bilinear products\footnote{this
  is a leading exponent of~\({\log_{2}{7}=\log_{4}{49}\approx{2.8074}}\)}
for~\(\matrixsize{2}{2}\) matrices requires~\(15\)~additional operations
yielding a leading term~\({6\,n^{\log_{2}\!{7}}}\) (\(12\) operations and
leading term~\({5\,n^{\log_{2}\!{7}}}\) for an alternative basis version).
This has been improved to require~\(156\) additional operations only in
the tensorized version for Strassen-Winograd algorithm for
\(\matrixsize{4}{4}\) matrices with~\(49\) bilinear products, and thus a
leading term~\({\left(5+\frac{8}{11}\right)n^{\log_{2}\!{7}}}\), see,
  e.g., the matrices and straight-line programs
  \plinoptdata{data/4x4x4_49_156_{L,R,P}.s{ms,lp}} in the
  \plinopt repository\footnote{\url{\plinoptgit}}.
Another comparison is with the~\(47\) bilinear products
algorithm\footnote{this is a leading exponent of~\({\log_{4}{47}\approx{2.7773}}\)}, working in characteristic~\(2\) only,
of~\cite{alphaevolve}: the best known straight-line program requires~\(170\) additional XORs, and thus a leading
term~\({\left(6+\frac{15}{31}\right)n^{\log_{4}\!{47}}}\), see, e.g.,
\plinoptdata{data/4x4x4_47_mod_2_{L,R,P}.s{ms,lp}} in the \plinopt
repository.
\par
We here present an algorithm for multiplying two~\(\matrixsize{4}{4}\)
matrices using~\(48\) bilinear multiplications that requires~\(204\)
additions, subtractions and scalings over any ring containing an
inverse of~\(2\).
This results in a recursive algorithm with a leading cost term of
\begin{equation}
	\left(1+\frac{204}{48-16}\right)n^{2+\log_{4}\!{3}}=\frac{59}{8}n^{2+\log_{4}\!{3}}=\left(7+\frac{3}{8}\right)n^{2+\log_{4}\!{3}}
	\approx{7.375\,n^{2.7925}}.
 \end{equation}
No external input or output basis conversion is required.
\par
We then construct an alternative basis version of this algorithm, with a leading term of
\begin{equation}
	\left(1+\frac{173}{48-16}\right)n^{2+\log_{4}\!{3}}=\frac{205}{32}n^{2+\log_{4}\!{3}}=\left(6+\frac{13}{32}\right)n^{2+\log_{4}\!{3}}\approx{6.40625\,n^{2.7925}}.
\end{equation}
This uses~\(173\) operations in the encoded residuum and~\(109\) in the
three conversions.
\par
An intermediate rectangular encoding with inner dimensions~\({d=24}\)
uses~\({122=32+32+58}\) residual operations and~\(151\) conversion
operations, giving a leading term:
\begin{equation}
 \left(1+\frac{122}{48-24}\right)n^{2+\log_{4}\!{3}}=\frac{73}{12}n^{2+\log_{4}\!{3}}=\left(6+\frac{1}{12}\right)n^{2+\log_{4}\!{3}}
	\approx{6.083\,n^{2.7925}}.
\end{equation}
Finally, we found a rectangular encoding with inner
dimensions~\({d=32}\) that reduces the third part to~\(32\) residual
operations and~\(63\) conversion operations, giving a leading term,
mixing the inner dimensions~\({d=24}\) and~\({d=32}\), of:
\begin{equation}
 \left(1+\frac{32}{48-24}+\frac{32}{48-24}+\frac{32}{48-32}\right)n^{2+\log_{4}\!{3}}=\frac{17}{3}n^{2+\log_{4}\!{3}}=\left(5+\frac{2}{3}\right)n^{2+\log_{4}\!{3}}
	\approx{5.667\,n^{2.7925}}.
\end{equation}
\subsection{Context: \FMMA{4}{4}{4}{48} algorithms}
An~\FMMA{m}{k}{n}{r} algorithm for multiplying an~\matrixsize{m}{k} matrix by a~\matrixsize{k}{n} matrix with~\(r\) multiplications can be described by these~\(r\) products and~\(3\) linear maps encoding the additions and scalings by constants.
\par
Consider a matrix multiplication~\({\mat{C}\leftarrow{\mat{A}\cdot{\mat{B}}}}\) and denote by~\(\vectorization{\mat{X}}\) the row-major vectorization of the matrix~\(\mat{X}\) in~\(\K^{\matrixsize{\firstdim}{\seconddim}}\), i.e.\ the vector~\(\vec{v}\) in~\(\K^{\firstdim\seconddim}\) of the matrix entries such that~\({{v_{i\seconddim+j} = a_{i,j}}}\) where indices~\({i,j}\) are~\(0\)-based.
\par
Letting~\(\odot\) denote the Hadamard~(entry-wise) product of vectors, a matrix multiplication algorithm is then linear combinations of the \emph{Left} input~\(\mat{A}\), of the \emph{Right} input~\(\mat{B}\), and of their \emph{Products} to form the output~\(\mat{C}\):
\begin{equation}
	\vectorization{\mat{C}}=\mat{P}\cdot\left(\mat{L}\cdot\vectorization{\mat{A}}\right)\odot(\mat{R}\cdot\vectorization{\mat{B}})\quad\text{with}\quad
	\mat{P}\ \text{in}\ \K^{k{\times}r},\ \mat{L}\ \text{in}\ \K^{r{\times}m}\ \text{and}\ \mat{R}\ \text{in}\ \K^{r{\times}n}.
\end{equation}
These~\({\mat{L},\mat{R}}\) and~\(\mat{P}\) matrices are called the \textsc{lrp} representation of the considered multiplication algorithm.
\par
In this setting, the rational algorithm for~\FMMA{4}{4}{4}{48} of~\cite{DPS25:444}
has an~\textsc{lrp} representation with matrices of dimensions \matrixsize{48}{16}, \matrixsize{48}{16} and \matrixsize{16}{48}.
Their respective number of non-zero~(including non-unit) coefficients
are~\({{448~(64)}, {288~(0)}}\) and~\({336~(304)}\), adding up to~\({1072~(368)}\).
Then the best known associated straight-line programs~(\SLPs) for these three matrices require respectively~\(104\)~additions and~\(4\)~scalings,~\(75\)~additions and one scaling~\(110\) additions and~\(21\) scalings.
This is a total of~\(315\) scalar operations and hence a dominant term of complexity bound~\({\frac{347}{32}n^{2+\log_{4}\!{3}}=\left(10+\frac{27}{32}\right)n^{2+\log_{4}\!{3}}\approx{10.84375n^{2.7925}}}\).
\par
In this note, we present different algorithms in the same de Groote isotropy orbit (see e.g.~\cite{DPS25:444} or references therein) that reduce this leading constant:
first, by sparsifying the~\textsc{lrp} representation and/or increasing the number of vectors of the canonical basis in this~\textsc{lrp} representation;
second, by optimizing the associated straight-line programs to improve on the number of required additional operations.
\subsection{Contributions: improving the time complexity bound leading constant}
In this note, we explore and combine several strategies used to
improve the fast matrix multiplication complexity bound constant:
\begin{itemize}
\item because the number of non-zero and non-unit coefficients in the
 algorithm impacts the arithmetic complexity, we seek an \textsc{lrp} representation that is as sparse as possible, as done in~\cite{Beniamini:2019aa,Beniamini:2020aa};
\item the analysis of existing representations (e.g.\ Winograd representation~\cite{Winograd:1977} (resp.~\cite{Karunaratne:2026aa}) of fast~\(\matrixsize{2}{2}\)-matrix multiplication (resp.~\(\matrixsize{3}{3}\)) have a better complexity exponent then Strassen representation~\cite{strassen:1969} (resp.\ Laderman~\cite{laderman:1976a})) shows that the number of canonical basis vectors as rows (resp.\ columns) of~\(\mat{L}\) and~\(\mat{R}\) (resp.~\(\mat{P}\)) also impacts the constant. We therefore aim to find an \textsc{lrp} representation that is as sparse as possible with maximizing the number of such vectors;
\item the technique of alternative basis introduced
 in~\cite{Karstadt:2017aa,Beniamini:2019aa} can also be used to
 reduce the arithmetic complexity constant.
\end{itemize}
We then combine these approaches to obtain the following results:
\begin{enumerate}
\item The sparsest variant we could find has~\({624=202*2+220}\)
 non-zero elements (among those,~\(104\) non-units).
 It can be realized by an \SLP with~\(230\) scalar operations, yielding
 a complexity bound constant of~\(8.1875\).
\item The variant with the largest number of vectors of the canonical
 basis we could find has~\(30\) of those, with~\(10\) such vectors
of the canonical bases in each one of~\({\mat{L},\mat{R}}\) and~\(\mat{P}\).
 It also has~\({760=256*2+248}\) non-zero elements (among
 those,~\(136\) non-units).
 It can be realized by an \SLP with~\(200\) additions/subtractions and~\(4\) halvings, improving the complexity bound constant to~\({\frac{59}{8}\approx{7.375}}\).
 Furthermore, symmetry between the~\(\mat{L}\) and the~\(\mat{R}\)
 allows to express the~\SLP encoding~\(\mat{R}\) to be expressed identically to
 the~\SLP encoding~\(\mat{L}\), modulo a permutation and whole
 row/column sign swaps on its
 inputs and outputs.
\item We then present alternative basis versions
 (see~\cite{Karstadt:2017aa,Beniamini:2019aa}) of these two variants
 yielding further improvements in the constants to respectively~\(6.78125\) and~\({\frac{103}{16}\approx{6.4375}}\).
A rectangular encoding of the canonical variant with inner dimension~\({d=24}\) further gives coefficient~\({\frac{73}{12}\approx{6.0833}}\).
\end{enumerate}
\Cref{tab:performance_comparison} summarizes the characteristics of the new algorithms presented in
this paper, including the (approximate) cost' leading constant when applied recursively.
\begin{table}[htbp]
\centering
\begin{tabular}{rrrrrrrr}
\toprule
 \multicolumn{2}{c}{Strategies} & Non-Zeros & Non-Units & Canonical & \(\#\pm\) & \(\#(\times/)\) & {Constant} \\
\midrule
\multicolumn{2}{c}{starting point~\cite{DPS25:444}} & 1072 & 368 & 0 & 289 & 26 & \(10.84\) \\
\midrule
\multirow{3}{*}{\begin{turn}{90}Sparse\end{turn}} & Regular & 624 & 104 & 4 & 225 & 5 & \({8.19}\) \\
& Rect.\ Alt.\ Basis & 149 & 0 & 141 & 6 & 0 & \({7.00}\) \\
& Square Alt.\ Basis& 418 & 106 & 48 & 177 & 8 & \({6.78}\) \\
\midrule
\multirow{5}{*}{\begin{turn}{90}Canonical\end{turn}}& Regular & 760 &136 & 30 & 200 & 4 & \(7.38\) \\
& Rect.\ Alt.\ Basis & 149 & 0 & 141 & 6 & 0 & \({7.00}\) \\
& Square Alt.\ Basis& 514 & 64 & 48 &171 & 2 & \(6.41\) \\
& Rect.\ \({d=24}\) & 434 & 38 & 64 &122 & 0 & \(6.08\) \\
& \({d=24}\) (\({\mat{L},\mat{R}}\)); \(32\)~(\(\mat{P}\))& 117+117+64 & 1+1+0 & 22+22+32 & 32+32+32 & 0 & \(5.67\) \\
\bottomrule
\end{tabular}
\caption{Characteristics of the variants of the \FMMA{4}{4}{4}{48}
  algorithms presented here. The leading constant~\(\lambda\) is
obtained from the number of additions, subtractions~(\({a=\#\pm}\))
and of scalings~(\({s=\#(\times/)}\)) as
  \({\lambda=1+\frac{a+s}{r-d}}\), where~\({r=48}\) is the tensor rank
  and~\(d\) in~\({\{16,24,32,47\}}\) is the column dimension of
  the~\({\mat{L},\mat{R}}\) and~\(\Transpose{\mat{P}}\) residual
  matrices (or with separated counts for each of the three~\({\mat{L},
    \mat{R}}\) and~\(\mat{P}\) if they have different inner dimension~\(d\)). Canonical
  vectors are counted up to sign. The alternative rows give residual
  costs; their conversion costs are accounted for
  separately.}\label{tab:performance_comparison}
\end{table}
%
\section{Sparse and canonical tensor decomposition}
Each fast matrix multiplication algorithm viewed as a bilinear map is associated with a tensor decomposition (see~\cite{Dumas:2026:autoaccurate} and references therein for more details on the notation and standpoint adopted in this note).
The \emph{de Groote isotropy group} is a natural group of automorphisms acting on these decompositions (see~\cite{groot:1978a});
it maps a tensor decomposition on another encoding an equivalent
algorithm (with same complexity exponent but potentially different other
properties, such as the arithmetic complexity constant).
\subsection{Sparse tensor decomposition in the de Groote orbit}\label{sec:sparse}
We identified a tensor decomposition in the de Groote orbit that ,to the best of our knowledge, features the sparsest \textsc{lpr} representation with only~\({{202~(32)}, {202~(32)}}\) and~\({220~(40)}\) non-zero elements
(including non-units), summing up to~\(624\)~(including~\(104\)
non-units).
The associated matrices are
presented in~\cref{eq:444LRP} below
and are
available as~\plinoptdata{data/4x4x4_48_sparse_{L,R,P}.sms} in
the~\plinopt repository.
\par
\begin{table}[ht]\centering
\resizebox{.9\linewidth}{!}{\(%
\mat{L}=\begin{smatrix}
1&0&1&0&0&0&0&0&1&0&1&0&0&0&0&0\\
0&0&0&0&0&0&0&2&0&0&0&1&0&0&0&-1\\
0&0&0&0&0&0&1&0&0&0&1&0&0&0&0&0\\
0&-1&0&1&0&0&0&0&\frac{1}{2}&0&0&0&0&0&0&0\\
0&0&0&0&0&0&0&0&0&0&0&0&0&1&1&0\\
0&0&0&1&0&0&0&-1&0&0&0&0&0&0&0&1\\
0&0&1&0&0&0&1&0&0&0&1&0&0&0&-1&0\\
0&0&0&0&1&0&0&-1&1&0&0&-1&0&0&0&0\\
0&\frac{1}{2}&0&0&0&-1&-2&0&0&0&-1&0&0&\frac{1}{2}&1&0\\
0&0&0&0&1&1&1&0&0&0&0&0&0&0&0&0\\
0&0&0&1&0&0&0&1&0&0&0&1&0&0&0&-1\\
0&\frac{1}{2}&0&0&1&0&0&-1&0&0&0&0&0&\frac{1}{2}&0&0\\
0&0&0&0&0&1&1&0&0&0&0&0&0&0&0&0\\
0&-1&0&1&0&0&0&0&0&0&0&0&0&0&0&0\\
0&0&0&0&0&0&1&0&\frac{1}{2}&0&1&0&0&0&0&0\\
0&0&0&0&-1&0&0&1&0&0&0&0&0&0&0&0\\
0&0&0&0&1&0&0&0&\frac{1}{2}&0&0&0&0&1&0&-1\\
0&-1&0&1&0&1&0&-1&0&0&0&0&0&-1&0&1\\
0&-\frac{1}{2}&0&1&0&0&0&-1&0&0&0&0&0&-\frac{1}{2}&0&1\\
0&1&0&-1&0&1&0&-1&0&1&0&-1&0&-1&0&1\\
0&0&0&1&0&0&0&0&0&0&0&1&0&0&0&0\\
0&-\frac{1}{2}&0&0&1&0&0&-1&1&0&0&-1&0&\frac{1}{2}&0&0\\
0&1&1&0&0&-1&-1&0&0&0&0&0&0&1&1&0\\
0&0&0&0&0&2&2&0&0&1&1&0&0&-1&-1&0\\
0&\frac{1}{2}&0&0&0&0&0&0&0&0&0&0&0&\frac{1}{2}&1&0\\
1&1&1&0&0&0&0&0&1&1&1&0&0&0&0&0\\
0&0&0&0&0&0&0&0&0&0&0&0&1&1&0&-1\\
1&\frac{1}{2}&1&0&0&-1&0&0&1&0&1&0&0&\frac{1}{2}&0&0\\
0&0&0&0&0&0&2&0&0&0&1&0&0&0&-1&0\\
1&1&0&-1&0&0&0&0&0&0&0&0&0&0&0&0\\
0&0&0&0&2&0&0&-2&1&0&0&-1&-1&0&0&1\\
0&-\frac{1}{2}&0&1&0&0&0&1&0&0&0&1&0&\frac{1}{2}&0&-1\\
0&0&0&0&0&0&0&0&0&0&0&0&0&0&1&0\\
0&0&0&1&0&0&0&0&-\frac{1}{2}&0&0&1&0&0&0&0\\
1&\frac{1}{2}&1&0&0&0&0&0&0&0&0&0&0&\frac{1}{2}&0&0\\
0&0&0&0&1&1&0&-1&1&1&0&-1&0&0&0&0\\
1&1&1&0&-1&-1&-1&0&0&0&0&0&1&1&1&0\\
1&1&1&0&0&-1&-1&0&\frac{1}{2}&0&0&0&0&1&1&0\\
1&0&1&0&1&0&1&0&1&0&1&0&-1&0&-1&0\\
0&0&0&0&1&0&0&-2&\frac{1}{2}&0&0&-1&0&0&0&1\\
0&0&0&0&1&1&0&-1&0&0&0&0&0&0&0&0\\
-1&0&0&1&0&0&0&0&-1&0&0&1&0&0&0&0\\
0&0&0&0&1&0&1&0&1&0&1&0&0&0&0&0\\
0&0&0&0&0&-1&-1&0&\frac{1}{2}&0&0&0&0&0&0&0\\
1&0&1&0&0&0&0&0&0&0&0&0&0&0&0&0\\
1&1&1&0&0&0&0&0&0&0&0&0&0&0&0&0\\
0&0&0&0&0&0&0&0&0&0&0&0&0&-1&0&1\\
1&0&1&0&0&0&1&0&\frac{1}{2}&0&1&0&0&0&-1&0\\
 \end{smatrix},\
\mat{R}=\begin{smatrix}
0&0&0&0&-1&0&0&-1&1&0&0&1&0&0&0&0\\
0&1&2&1&0&0&0&0&0&0&0&0&0&1&2&1\\
-1&1&1&0&0&0&0&0&1&-1&-1&0&-1&1&1&0\\
1&0&0&0&0&0&0&0&-1&0&0&0&\frac{1}{2}&-\frac{1}{2}&0&0\\
0&0&0&0&0&1&0&0&0&0&0&0&0&1&0&0\\
0&0&0&0&0&0&1&1&0&0&-1&-1&0&0&1&1\\
-1&-1&-1&-1&0&0&0&0&1&1&1&1&0&0&0&0\\
0&0&0&0&-1&1&1&0&0&0&0&0&-1&1&1&0\\
0&0&0&0&0&1&2&1&0&0&-1&-\frac{1}{2}&0&0&0&0\\
0&0&1&0&0&0&0&0&0&0&0&0&0&0&1&0\\
0&0&0&0&1&1&1&1&0&0&0&0&1&1&1&1\\
0&0&0&0&0&0&1&0&0&0&-1&0&\frac{1}{2}&-\frac{1}{2}&0&0\\
0&0&1&0&0&0&0&0&0&0&-1&0&0&0&1&0\\
1&0&0&0&0&0&0&0&-1&0&0&0&1&0&0&0\\
-1&1&1&0&0&0&0&0&1&0&0&\frac{1}{2}&-1&1&1&0\\
0&0&0&0&0&0&1&0&0&0&-1&0&0&0&1&0\\
0&1&0&0&0&0&0&0&0&0&0&0&-\frac{1}{2}&\frac{1}{2}&0&0\\
0&0&0&0&0&0&1&1&0&0&-1&-1&0&0&0&0\\
0&0&0&0&0&0&-1&-1&0&0&1&1&\frac{1}{2}&\frac{1}{2}&0&0\\
0&0&0&0&1&1&1&1&0&0&0&0&0&0&0&0\\
1&0&0&1&0&0&0&0&-1&0&0&-1&1&0&0&1\\
0&0&0&0&-1&1&1&0&0&0&0&0&-\frac{1}{2}&\frac{1}{2}&0&0\\
0&0&-1&-1&0&0&0&0&0&0&1&1&0&0&0&0\\
0&0&0&0&0&1&2&1&0&0&0&0&0&0&0&0\\
0&0&0&0&0&1&0&0&0&0&1&\frac{1}{2}&0&1&0&0\\
0&0&0&0&1&0&0&1&0&0&0&0&0&0&0&0\\
0&1&0&0&0&0&0&0&0&0&0&0&0&0&0&0\\
0&0&0&0&1&0&0&1&0&0&0&-\frac{1}{2}&0&0&0&0\\
0&0&0&0&0&1&2&1&0&-1&-2&-1&0&0&0&0\\
-1&0&0&0&0&0&0&0&1&0&0&0&0&0&0&0\\
0&1&2&1&0&0&0&0&0&0&0&0&0&0&0&0\\
0&0&0&0&1&1&1&1&0&0&0&0&\frac{1}{2}&\frac{1}{2}&0&0\\
0&0&0&0&0&-1&0&0&0&1&0&0&0&-1&0&0\\
-1&0&0&-1&0&0&0&0&1&0&0&1&-\frac{1}{2}&\frac{1}{2}&1&0\\
0&0&0&0&1&0&0&0&0&0&0&\frac{1}{2}&1&0&0&0\\
0&0&0&0&-1&1&1&0&0&0&0&0&0&0&0&0\\
0&0&1&1&0&0&0&0&0&0&0&0&0&0&0&0\\
0&0&1&1&0&0&0&0&0&0&0&-\frac{1}{2}&0&0&0&0\\
1&1&1&1&0&0&0&0&0&0&0&0&0&0&0&0\\
0&1&2&1&0&0&0&0&0&0&0&0&-\frac{1}{2}&\frac{1}{2}&1&0\\
0&0&0&0&0&0&-1&0&0&0&1&0&0&0&0&0\\
-1&0&0&-1&0&0&0&0&1&0&0&1&0&0&0&0\\
1&-1&-1&0&0&0&0&0&0&0&0&0&1&-1&-1&0\\
0&0&1&0&0&0&0&0&0&0&0&\frac{1}{2}&0&0&1&0\\
0&0&0&0&-1&0&0&0&1&0&0&0&-1&0&0&0\\
0&0&0&0&1&0&0&0&0&0&0&0&1&0&0&0\\
0&1&0&0&0&0&0&0&0&0&0&0&0&1&0&0\\
1&1&1&1&0&0&0&0&-1&0&0&-\frac{1}{2}&0&0&0&0\\
\end{smatrix}\)}
\resizebox{.95\linewidth}{!}{\(%
\mat{P}=\begin{smatrix}
0&0&0&0&0&0&0&0&0&0&0&0&0&1&0&0&0&0&0&0&0&0&0&0&0&0&0&0&0&-1&0&0&0&0&0&0&0&0&0&0&0&0&0&0&1&1&0&0\\
0&0&1&-2&-1&0&1&0&0&0&0&0&1&1&-1&0&0&0&0&0&0&0&-1&0&2&0&0&0&0&1&0&0&1&0&0&0&0&-1&0&0&0&0&0&1&0&0&0&1\\
0&0&0&1&1&0&0&1&0&0&0&1&-1&0&0&1&0&0&0&0&0&-1&1&0&-1&0&0&0&0&-1&0&0&0&1&-1&0&0&1&0&0&0&-1&0&-1&0&1&0&0\\
0&0&0&0&0&1&0&-1&0&0&1&-1&0&-1&0&-1&0&0&1&0&-1&1&0&0&0&0&0&0&0&1&0&-1&0&-2&2&0&0&0&0&0&0&1&0&0&0&-2&0&0\\
-1&0&0&-1&0&0&0&0&0&1&0&-1&0&1&1&0&1&-1&-1&0&0&0&0&0&0&0&0&-1&0&0&0&0&0&0&1&0&0&0&0&0&-1&0&1&1&1&0&1&0\\
0&0&1&-1&0&0&0&0&-1&0&0&1&1&1&-1&0&1&-1&-1&0&0&0&0&0&1&0&0&0&-1&0&0&0&1&0&0&0&0&0&0&0&1&0&0&1&0&0&1&0\\
0&0&0&0&0&0&0&0&0&1&0&0&-1&0&0&1&0&0&0&0&0&0&0&0&0&0&0&0&0&0&0&0&0&0&0&0&0&0&0&0&-1&0&0&0&0&0&0&0\\
0&1&0&1&0&1&0&0&0&-2&0&0&0&-1&0&-1&-1&1&2&0&-1&0&0&0&0&0&0&0&0&0&0&0&0&-1&0&0&0&0&0&1&1&0&0&-2&0&0&-1&0\\
1&0&0&2&0&0&0&0&0&0&0&1&0&-2&0&0&0&1&1&1&0&1&0&0&0&-1&0&2&0&0&0&1&0&0&0&-1&0&0&0&0&1&0&0&0&-1&-1&0&0\\
0&0&-2&2&1&0&0&0&0&0&0&-1&-2&-2&2&0&0&1&1&1&0&-1&0&-1&-2&0&0&0&1&0&0&1&-1&0&0&1&0&0&0&0&-1&0&0&-2&0&0&0&0\\
0&0&0&0&-1&0&0&-1&1&0&0&0&2&0&0&-1&0&0&0&0&0&2&0&1&1&-1&0&1&0&0&0&0&0&0&1&-1&0&0&0&0&1&0&0&2&0&-1&0&0\\
0&0&0&-2&0&-1&0&1&0&0&-1&0&0&2&0&1&0&-1&-2&-1&2&-2&0&0&0&2&0&-2&0&0&0&0&0&2&-2&1&0&0&0&0&-1&0&0&0&0&2&0&0\\
0&0&0&0&0&0&0&0&0&1&0&0&0&0&1&0&2&0&0&0&0&0&0&0&0&0&-1&0&0&0&0&0&0&0&2&0&-1&1&-1&0&0&0&1&1&1&-1&1&1\\
0&0&0&0&1&0&0&0&0&0&0&0&0&0&0&0&0&0&0&0&0&0&0&0&0&0&1&0&0&0&0&0&1&0&0&0&0&0&0&0&0&0&0&0&0&0&1&0\\
0&0&0&0&-1&0&0&-1&0&1&0&1&0&0&0&1&1&0&0&0&0&1&0&0&1&0&-1&0&0&0&-1&0&0&0&1&0&-1&1&0&1&0&0&0&1&0&-1&0&0\\
0&1&0&0&0&1&0&1&0&-2&-1&-1&0&0&0&-1&-2&0&1&0&0&-1&0&0&0&0&1&0&0&0&1&1&0&0&-2&0&2&-2&0&0&0&0&0&-2&0&2&-1&0\\
\end{smatrix}
\)}
\caption{Sparsest known \FMMA{4}{4}{4}{48} \textsc{lpr} representation}\label{eq:444LRP}
\end{table}
\par
Note that the obtained matrices~\(\mat{L}\) and~\(\mat{R}\) are in fact identical
up to row/column permutations and whole row/column sign swaps.
\subsection{Maximizing the number of vectors from the canonical basis}\label{sec:canonical}
A program realizing the matrix vector product by~\(\mat{L}\) or~\(\mat{R}\)
(resp.~\(\mat{P}\)), has to produce the output of all the dot-products by
their rows (resp.\ columns). A direct lower bound for the number of
operations required to perform these operations can then be obtained
by looking at these rows (resp.\ columns): any row (resp.\ column) of
the~\(\mat{L}\) and~\(\mat{R}\) (resp.~\(\mat{P}\)) matrices that is not a vector of the
canonical basis (up to sign), and distinct from the other rows
(resp.\ columns), requires at least one operation to be computed, even
from previously computed intermediate values.
\par
From this, the smallest found lower bound (on programs realizing the matrix multiplication within the de Groote orbit) is obtained by maximizing the number of vectors from the canonical basis within the~\({\mat{L},\mat{R}}\) and~\(\mat{P}\) matrices.
\par
In the same orbit, we were able to find a tensor decomposition whose associated~\textsc{lrp} representation has~\({{256~(64)},{256~(64)}}\) and~\({248~(8)}\) non-zero elements, summing up to~\(760\)~(including~\(136\) non-units), but with~\(10\) vectors of the canonical basis as rows (resp.\ columns) of each one of the~\(\mat{L}\) and~\(\mat{R}\) (resp.~\(\mat{P}\)) matrices, summing up to~\(30\) vectors of the canonical basis overall.
\par
The associated matrices are available as
\plinoptdata{data/4x4x4_48_204_{L,R,P}.sms} in the \plinopt repository
and presented in~\cref{eq:canLRP} hereafter.
\par
\begin{table}[ht]\centering
\resizebox{.9\linewidth}{!}{\(%
\mat{L}=\begin{smatrix}
1&1&0&1&0&0&0&0&0&0&0&0&0&0&0&0\\
0&0&0&0&0&1&1&-1&0&0&0&0&0&0&0&0\\
0&0&0&0&0&0&0&0&-1&0&1&-1&0&0&0&0\\
1&1&1&0&0&0&0&0&-1&-\frac{1}{2}&-\frac{1}{2}&0&1&\frac{1}{2}&\frac{1}{2}&0\\
0&0&0&0&0&0&-1&0&0&0&1&0&0&0&1&0\\
0&-1&-1&1&0&1&1&-1&0&0&0&0&0&-1&-1&1\\
-1&0&1&-1&-1&0&1&-1&1&0&-1&1&0&0&0&0\\
0&0&0&0&0&0&0&0&0&0&0&1&0&0&0&0\\
\frac{1}{2}&0&0&\frac{1}{2}&\frac{1}{2}&0&-1&\frac{1}{2}&0&0&0&0&0&0&0&0\\
0&0&0&0&0&0&0&0&0&0&0&0&0&1&0&0\\
0&1&1&-1&0&1&1&-1&0&-1&-1&1&0&0&0&0\\
\frac{1}{2}&0&0&\frac{1}{2}&-\frac{1}{2}&0&0&-\frac{1}{2}&0&0&0&0&1&0&0&0\\
0&0&0&0&0&0&0&0&0&0&0&0&0&0&1&0\\
-1&-1&-1&0&0&0&0&0&1&1&1&0&-1&-1&-1&0\\
0&0&0&0&0&0&0&0&1&\frac{1}{2}&-\frac{1}{2}&1&0&-\frac{1}{2}&-\frac{1}{2}&0\\
0&0&0&0&0&0&0&0&0&0&0&0&0&0&0&1\\
0&0&0&0&-1&-1&-1&0&1&\frac{1}{2}&\frac{1}{2}&0&1&\frac{1}{2}&\frac{1}{2}&0\\
1&1&1&0&-1&-1&-1&0&0&0&0&0&1&1&1&0\\
\frac{1}{2}&1&1&-\frac{1}{2}&-\frac{1}{2}&-1&-1&\frac{1}{2}&0&0&0&0&1&1&1&0\\
1&1&1&0&1&1&1&0&-1&-1&-1&0&0&0&0&0\\
0&1&1&-1&0&0&0&0&0&0&0&0&0&0&0&0\\
\frac{1}{2}&0&0&\frac{1}{2}&\frac{1}{2}&0&0&\frac{1}{2}&-1&0&0&0&0&0&0&0\\
0&0&-1&0&0&0&1&0&0&0&0&0&0&0&-1&0\\
0&0&0&0&0&0&1&0&0&0&0&0&0&0&0&0\\
\frac{1}{2}&0&0&\frac{1}{2}&\frac{1}{2}&0&-1&\frac{1}{2}&-1&0&1&-1&0&0&1&0\\
0&1&0&0&0&0&0&0&0&0&0&0&0&0&0&0\\
0&0&0&0&-1&0&0&0&1&0&0&0&1&0&0&0\\
\frac{1}{2}&1&0&\frac{1}{2}&\frac{1}{2}&0&0&\frac{1}{2}&0&0&0&0&0&0&0&0\\
0&0&0&0&1&0&-1&1&0&0&0&0&0&0&0&0\\
-1&0&0&0&0&0&0&0&1&0&0&0&-1&0&0&0\\
0&0&0&0&0&0&0&1&0&0&0&0&0&0&0&0\\
-\frac{1}{2}&-1&-1&\frac{1}{2}&-\frac{1}{2}&-1&-1&\frac{1}{2}&1&1&1&0&0&0&0&0\\
0&0&0&0&-1&0&1&-1&1&0&-1&1&1&0&-1&1\\
0&-1&-1&1&0&0&0&0&0&\frac{1}{2}&\frac{1}{2}&0&0&-\frac{1}{2}&-\frac{1}{2}&0\\
\frac{1}{2}&1&0&\frac{1}{2}&\frac{1}{2}&0&0&\frac{1}{2}&-1&-1&0&-1&0&1&0&0\\
0&0&0&0&0&0&0&0&1&0&0&0&0&0&0&0\\
0&1&0&0&0&-1&0&0&0&0&0&0&0&1&0&0\\
0&1&0&0&0&0&1&0&0&-\frac{1}{2}&-\frac{1}{2}&0&0&\frac{1}{2}&-\frac{1}{2}&0\\
1&1&0&1&1&1&0&1&-1&-1&0&-1&0&0&0&0\\
0&0&0&0&0&-1&-1&1&0&\frac{1}{2}&\frac{1}{2}&0&0&\frac{1}{2}&\frac{1}{2}&0\\
0&0&0&0&0&0&0&0&0&0&0&0&1&0&0&0\\
0&0&0&1&0&0&0&0&0&0&0&0&0&0&0&0\\
0&0&0&0&0&0&0&0&1&1&0&1&0&0&0&0\\
0&0&0&0&0&0&0&0&0&\frac{1}{2}&\frac{1}{2}&0&0&-\frac{1}{2}&\frac{1}{2}&0\\
1&1&0&1&0&0&0&0&-1&-1&0&-1&1&1&0&1\\
0&1&0&0&0&0&0&0&0&-1&0&0&0&1&0&0\\
0&0&0&0&-1&-1&-1&0&1&1&1&0&1&1&1&0\\
1&1&0&1&1&0&-1&1&-1&-\frac{1}{2}&\frac{1}{2}&-1&0&\frac{1}{2}&\frac{1}{2}&0\\
 \end{smatrix},\
\mat{R}=\begin{smatrix}
1&1&0&-1&0&0&0&0&0&0&0&0&0&0&0&0\\
0&0&0&0&0&1&1&1&0&0&0&0&0&0&0&0\\
0&0&0&0&0&0&0&0&-1&0&1&1&0&0&0&0\\
\frac{1}{2}&0&\frac{1}{2}&0&0&0&0&0&-1&0&0&0&-\frac{1}{2}&0&-\frac{1}{2}&0\\
0&0&0&0&0&0&-1&0&0&0&1&0&0&0&1&0\\
0&0&0&0&0&0&0&0&0&0&0&0&0&1&0&0\\
-1&-1&-1&0&0&0&0&0&1&1&1&0&1&1&1&0\\
0&0&0&0&1&0&-1&-1&-1&0&1&1&-1&0&1&1\\
0&1&1&1&0&-\frac{1}{2}&-1&-\frac{1}{2}&0&\frac{1}{2}&1&\frac{1}{2}&0&0&0&0\\
0&0&0&0&0&0&0&1&0&0&0&0&0&0&0&0\\
0&0&0&0&-1&-1&-1&0&1&1&1&0&1&1&1&0\\
-\frac{1}{2}&0&\frac{1}{2}&1&0&0&0&0&0&0&0&0&\frac{1}{2}&0&-\frac{1}{2}&0\\
0&0&0&0&0&0&0&0&0&0&0&1&0&0&0&0\\
0&0&0&0&0&0&0&0&1&0&0&0&0&0&0&0\\
0&0&0&0&0&\frac{1}{2}&1&\frac{1}{2}&-1&-\frac{1}{2}&0&\frac{1}{2}&0&0&0&0\\
0&0&0&0&0&0&0&0&0&0&0&0&0&0&0&1\\
\frac{1}{2}&0&\frac{1}{2}&0&0&0&-1&0&0&0&0&0&-\frac{1}{2}&0&-\frac{1}{2}&0\\
0&1&0&0&0&0&0&0&0&0&0&0&0&0&0&0\\
\frac{1}{2}&1&\frac{1}{2}&0&0&0&0&0&0&0&0&0&-\frac{1}{2}&0&-\frac{1}{2}&0\\
1&1&1&0&-1&-1&-1&0&1&1&1&0&0&0&0&0\\
0&0&0&0&0&0&0&0&1&1&0&-1&0&0&0&0\\
-\frac{1}{2}&0&\frac{1}{2}&1&1&0&-1&-1&-1&0&1&1&-\frac{1}{2}&0&\frac{1}{2}&0\\
0&1&0&0&0&0&0&0&0&-1&0&0&0&-1&0&0\\
0&1&1&1&0&-1&-1&-1&0&1&1&1&0&0&0&0\\
0&0&0&0&0&-\frac{1}{2}&-1&-\frac{1}{2}&0&\frac{1}{2}&1&\frac{1}{2}&0&0&1&0\\
1&1&0&-1&-1&-1&0&1&1&1&0&-1&0&0&0&0\\
0&0&-1&0&0&0&1&0&0&0&0&0&0&0&1&0\\
1&1&0&-1&-1&-\frac{1}{2}&0&\frac{1}{2}&1&\frac{1}{2}&0&-\frac{1}{2}&0&0&0&0\\
0&1&1&1&0&0&0&0&0&0&0&0&0&0&0&0\\
1&0&0&0&0&0&0&0&-1&0&0&0&-1&0&0&0\\
0&-1&-1&-1&0&1&1&1&0&0&0&0&0&1&1&1\\
\frac{1}{2}&1&\frac{1}{2}&0&-1&-1&-1&0&1&1&1&0&\frac{1}{2}&0&\frac{1}{2}&0\\
0&0&0&0&0&0&0&0&0&0&0&0&0&0&1&0\\
\frac{1}{2}&1&\frac{1}{2}&0&0&0&0&0&-1&-1&0&1&-\frac{1}{2}&-1&-\frac{1}{2}&0\\
0&0&0&0&1&\frac{1}{2}&0&-\frac{1}{2}&-1&-\frac{1}{2}&0&\frac{1}{2}&-1&0&0&0\\
1&0&-1&-1&-1&0&1&1&1&0&-1&-1&0&0&0&0\\
0&-1&0&0&0&1&0&0&0&0&0&0&0&1&0&0\\
0&-1&0&0&0&\frac{1}{2}&0&\frac{1}{2}&0&\frac{1}{2}&0&-\frac{1}{2}&0&1&0&0\\
-1&-1&-1&0&1&1&1&0&0&0&0&0&1&1&1&0\\
-\frac{1}{2}&-1&-\frac{1}{2}&0&0&1&1&1&0&0&0&0&\frac{1}{2}&1&\frac{1}{2}&0\\
0&0&0&1&0&0&0&0&0&0&0&0&0&0&0&0\\
-1&-1&0&1&0&0&0&0&1&1&0&-1&1&1&0&-1\\
0&0&0&0&1&0&-1&-1&0&0&0&0&0&0&0&0\\
0&0&0&0&0&\frac{1}{2}&0&\frac{1}{2}&0&-\frac{1}{2}&0&\frac{1}{2}&0&0&0&0\\
0&0&0&0&0&0&0&0&0&0&0&0&1&0&0&0\\
0&0&0&0&-1&0&0&0&1&0&0&0&1&0&0&0\\
0&0&0&0&0&0&1&0&0&0&0&0&0&0&0&0\\
1&1&1&0&0&-\frac{1}{2}&-1&-\frac{1}{2}&-1&-\frac{1}{2}&0&\frac{1}{2}&-1&-1&-1&0\\
\end{smatrix}\)}
\par
\resizebox{.95\linewidth}{!}{\(%
\mat{P}=\begin{smatrix}
1&0&0&-2&0&0&0&0&0&0&0&1&0&1&0&0&0&-1&1&1&0&-1&0&0&0&1&0&-2&0&-1&0&1&0&0&0&1&0&0&0&0&-1&0&0&0&0&0&0&0\\
0&0&0&1&0&0&0&0&1&0&0&0&-1&-1&0&0&0&1&-1&-1&1&0&1&1&0&-1&0&1&0&0&0&-1&0&-1&0&0&0&1&0&0&0&0&0&1&0&0&0&0\\
0&0&1&0&0&0&1&0&0&0&0&-1&1&1&1&0&0&-1&1&1&0&1&-1&-1&0&0&0&0&-1&1&0&1&0&0&0&-1&0&-1&0&0&1&0&0&-1&0&0&0&-1\\
0&0&0&-1&0&0&0&0&1&0&0&1&-1&0&0&0&0&0&0&0&0&-1&1&1&0&1&0&-1&0&-1&0&0&0&-1&0&1&0&1&0&0&-1&-1&0&1&0&0&0&0\\
-1&0&0&0&0&0&0&0&0&1&0&-1&0&0&1&0&0&1&-1&1&0&-1&0&0&0&1&-1&0&0&0&0&1&0&0&0&1&1&-1&1&0&1&0&1&1&0&0&1&1\\
0&1&0&0&0&0&0&0&1&-1&0&0&0&0&0&0&-1&-1&1&-1&0&0&0&1&0&-1&0&1&0&0&0&-1&0&0&0&0&-1&1&0&1&0&0&0&-1&0&0&-1&0\\
0&0&0&0&0&0&0&0&-2&0&0&1&0&0&0&0&2&1&-1&1&0&1&0&-1&0&0&1&0&1&0&0&1&0&0&0&-1&0&0&0&0&-1&0&0&0&0&0&1&0\\
0&0&0&0&0&0&0&0&1&1&0&-1&0&0&0&0&-1&0&0&0&0&-1&0&1&0&1&-1&-1&0&0&1&0&0&0&0&1&1&-1&0&-1&1&0&0&1&0&0&0&0\\
0&0&0&-1&0&0&0&0&0&1&0&0&0&1&1&0&1&0&0&1&0&-1&0&0&0&1&0&-1&0&0&0&1&0&0&1&1&0&0&0&0&0&0&1&1&0&1&1&0\\
0&1&0&1&-1&0&0&0&1&-1&1&0&-1&-1&0&0&-1&0&0&-1&1&0&0&1&1&-1&0&1&0&0&0&0&0&-1&-1&0&0&0&0&1&0&0&0&0&0&-1&-1&0\\
0&0&1&-1&1&0&0&0&-1&0&0&0&1&1&1&0&1&0&0&1&0&1&0&-1&-1&0&0&0&0&0&0&1&0&0&0&-1&0&0&0&0&0&0&0&-1&0&0&1&0\\
0&0&0&0&-1&0&0&1&1&1&0&0&-1&0&0&0&0&0&0&0&0&-2&0&1&1&1&0&-1&0&0&0&0&0&0&1&1&0&0&0&0&0&0&0&2&0&1&0&0\\
-1&0&0&1&0&0&0&0&0&1&0&-1&0&-1&1&0&1&1&-1&0&0&0&0&0&0&0&0&1&0&0&0&0&0&0&1&0&0&0&0&0&1&0&1&1&1&0&1&0\\
0&1&0&-1&0&1&0&0&0&-1&0&0&1&1&0&0&-1&-1&2&0&-1&0&0&0&0&0&0&0&0&0&0&0&0&1&0&0&0&0&0&1&0&0&0&-2&0&0&-1&0\\
0&0&-1&1&0&0&0&0&-1&0&0&1&-1&-1&-1&0&1&1&-1&0&0&0&0&0&1&0&0&0&1&0&0&0&1&0&0&0&0&0&0&0&-1&0&0&1&0&0&1&0\\
0&0&0&0&0&0&0&0&0&1&0&0&1&0&0&1&0&0&0&0&0&0&0&0&0&0&0&0&0&0&0&0&0&0&0&0&0&0&0&0&1&0&0&0&0&0&0&0\\
\end{smatrix}
\)}
\caption{\FMMA{4}{4}{4}{48} algorithm with~\(30\) vectors of the canonical basis}\label{eq:canLRP}
\end{table}
Again, the obtained matrices~\(\mat{L}\) and~\(\mat{R}\) are in fact
identical up to row/column permutations and whole row/column sign
swaps.
\section{Short straight-line programs}
We have developed the \plinopt library that can produce short straight-line programs for linear programs.
Since finding the optimal \SLP is~\textsc{np}-hard~\cite{Boyar:2008:shortestSLP,Morgenstern:1975:Linear}, the \texttt{optimizer} strategy of this library combines several
heuristics to produce an \SLP from the matrix of a linear application:
\begin{itemize}
\item Random walks of common subexpression elimination (\textsc{cse}, see e.g.~\cite{Perminov:2025:parallelCSE, Martensson:2026ab} and references therein);
\item Kernel methods expressing parts of the matrix as linear combinations of the other parts~\cite{jgd:2024:plinopt,Dumas:2026:autoaccurate,Dumas:2026:addsf}.
\end{itemize}
The \texttt{mirabelle} and \texttt{chartreuse} strategies of \plinopt then refine any \SLP using combinations of the following techniques:
\begin{itemize}
\item For \texttt{chartreuse.sh}: Finding linear dependencies between the variables of the \SLP (that is, now including the temporary variables);
\item For \texttt{mirabelle.sh}: Applying the \texttt{optimizer}
  strategy on a sub-program of the~\SLP (with \texttt{prune.sh}
  applying \texttt{mirabelle.sh} successively on all sub-programs
  involving a given number of temporary variables).
\end{itemize}
We have applied all these techniques on the variants of~\cref{sec:sparse,sec:canonical}.
\par
The shortest programs reported here were then obtained by a complementary search, whose details are left to an extended version.
In short, each intermediate value of an existing \SLP is viewed as its vector of coefficients; a few new intermediate values, sums or differences of existing ones, are proposed; and the program is rebuilt from its inputs using only the values it needs.
The program for~\(\mat{R}\) is then deduced from that for~\(\mat{L}\) by their signed equivalence, and the program for~\(\mat{P}\) is also optimized through its transpose.
For the alternative basis variants, the change of basis programs are optimized in the same way.
All programs and their operation counts are verified by exact replay.
\par
In the following, the leading constants are derived from recurrent relations using the following variant of the Master
Theorem.
\begin{lemma}\label{lem:master-thm}
Consider the non-negative recurrence relation
\({T(n) = a \cdot T\!\left(\frac{n}{b}\right) + c\cdot{\!\left(\frac{n}{b}\right)}^\beta}\)
and let~\(\alpha\) be~\({\log_{b}{a}}\).
 \begin{enumerate}
 \item If~\({\beta < \alpha}\), then~\({T(n)\leq T(1) n^{\alpha }+\frac{c}{{a-b^\beta}}n^\alpha}\).
 \item If~\({\beta > \alpha}\), then~\({T(n)\leq T(1)n^\alpha + \frac{c}{b^\beta-a}n^\beta}\).
 \end{enumerate}
 \end{lemma}
\subsection{SLPs for the sparse LRP representation}
The straight-line programs obtained from the sparsest~\({\mat{L},\mat{R},\mat{P}}\) matrices presented in~\cref{eq:444LRP} are available as \plinoptdata{data/4x4x4_48_sparse_{L,R,P}.slp} in the \plinopt repository.
\par
These straight-line programs require:
\begin{itemize}
\item~\(63\) additions and two scalings (binary shifts) for~\(\mat{L}\);
\item~\(63\) additions and two scalings (binary shifts) for~\(\mat{R}\);
\item~\(99\) additions and one scaling (binary shift) for~\(\mat{P}\).
\end{itemize}
This gives a total of~\(230\) operations and a theoretical complexity bound dominated by:
\begin{equation}
\left(1+\frac{230}{48-16}\right)n^{2+\log_4{\!3}}\approx{8.1875n^{2.7925}}.
\end{equation}
\subsection{SLPs for the canonical LRP representation}
For the canonical matrices of~\cref{eq:canLRP}, the improved
programs are given
in~\cref{lst:canL,lst:canR,lst:canH,lst:canP}. Their matching \SLP and
\textsc{sms} files,
are available as~\plinoptdata{data/4x4x4_48_204_{L,R,P}.s{ms,lp}} in
the \plinopt repository.

The programs act directly on the ordinary entries of~\({\mat{A},\mat{B}}\) and produce~\(\mat{C}\), with no input or output basis conversion. Each binary addition/subtraction, non-unit scaling and explicit unary negation is counted; the displayed programs have no unary negations. Copies and wire permutations cost zero.
\par
\begin{lstlisting}[style=slp,caption={\SLP for the input forms~\({\mat{L}}\)},label=lst:canL]
x10:=A[1,1]+A[1,4]; x11:=A[1,2]+A[2,3]; x14:=A[2,1]+A[2,4]; x21:=A[3,1]+A[3,4]; x25:=A[1,2]+A[4,2]; x26:=A[2,3]-A[4,3]; l0:=A[1,2]+x10; l28:=x14-A[2,3]; l2:=A[3,3]-x21; l42:=A[3,2]+x21; l36:=x25-A[2,2]; l45:=x25-A[3,2]; l22:=x26-A[1,3]; l4:=A[3,3]-x26; x12:=(x10+x14)/2; x13:=l4-l45; l27:=A[1,2]+x12; l21:=x12-A[3,1]; l8:=x12-A[2,3]; x15:=x10-x12; l11:=A[4,1]+x15; l43:=(x11+x13)/2; x16:=A[4,4]+l11; l37:=x11-l43; x17:=l43-A[4,3]; x18:=A[4,2]+l43; x19:=l8+l43; x20:=l37-l36; x22:=l37-l22; l14:=x17-l2; l39:=A[2,4]-x20; l33:=A[1,4]-x22; l24:=x19-l14; l1:=x18-l39; x23:=l11+l39; l20:=x17-l33; x24:=l39+l33; l34:=l24-x13; l32:=x16-l24; l16:=x23-l21; l18:=x23-l33; l10:=A[3,4]-x24; l31:=x24-l21; l44:=x16+l34; l47:=x19+l34; l26:=x20+l16; l46:=x18+l16; l17:=x15+l18; l5:=x16-l18; l3:=x23-l31; l19:=x12-l31; l38:=x20+l47; l6:=x22-l47; l29:=x22-l3; l13:=x17-l3; l7:=A[3,4]; l9:=A[4,2]; l12:=A[4,3]; l15:=A[4,4]; l23:=A[2,3]; l25:=A[1,2]; l30:=A[2,4]; l35:=A[3,1]; l40:=A[4,1]; l41:=A[1,4];
\end{lstlisting}
\begin{lstlisting}[style=slp,caption={\SLP for the input forms~\({\mat{R}}\)},label=lst:canR]
y10:=B[1,3]-B[4,3]; y11:=B[2,3]-B[3,1]; y14:=B[1,1]-B[4,1]; y21:=B[1,2]-B[4,2]; y25:=B[2,3]+B[2,4]; y26:=B[3,1]-B[3,4]; r26:=B[2,3]-y10; r29:=y14-B[3,1]; r22:=y21-B[3,2]; r36:=B[2,2]-y21; r42:=B[2,1]-y25; r1:=B[2,2]+y25; r2:=B[3,3]-y26; r20:=B[3,2]+y26; y12:=(y10+y14)/2; y13:=r1-r20; r16:=y12-B[2,3]; r18:=B[1,2]+y12; r3:=y12-B[3,1]; y15:=y12-y10; r11:=B[1,4]-y15; r43:=(y13-y11)/2; y16:=r11-B[4,4]; r14:=y11+r43; y17:=r43-B[3,4]; y18:=r43-B[2,4]; y19:=r3+r43; y20:=r42+r14; y22:=r2-r14; r37:=y17-r22; r34:=y20-B[4,1]; r24:=B[4,3]+y22; r33:=y19-r37; r45:=y18-r34; y23:=r11+r34; r4:=r24+y17; y24:=r24-r34; r39:=y13-r33; r41:=y16-r33; r27:=r18-y23; r21:=r24+y23; r10:=B[4,2]+y24; r31:=r18+y24; r30:=r39-y16; r47:=y19-r39; r0:=y20+r27; r25:=r27-y18; r35:=y15-r21; r7:=r21-y16; r8:=y23+r31; r19:=y12+r31; r38:=y20-r47; r6:=y22-r47; r28:=r8-y22; r23:=r8-y17; r5:=B[4,2]; r9:=B[2,4]; r12:=B[3,4]; r13:=B[3,1]; r15:=B[4,4]; r17:=B[1,2]; r32:=B[4,3]; r40:=B[1,4]; r44:=B[4,1]; r46:=B[2,3];
\end{lstlisting}
\par\medskip
\begin{lstlisting}[style=slp, caption={Products},label=lst:canH]
p0:=l0*r0; p1:=l1*r1; p2:=l2*r2; p3:=l3*r3; p4:=l4*r4; p5:=l5*r5; p6:=l6*r6; p7:=l7*r7; p8:=l8*r8; p9:=l9*r9; p10:=l10*r10; p11:=l11*r11; p12:=l12*r12; p13:=l13*r13; p14:=l14*r14; p15:=l15*r15; p16:=l16*r16; p17:=l17*r17; p18:=l18*r18; p19:=l19*r19; p20:=l20*r20; p21:=l21*r21; p22:=l22*r22; p23:=l23*r23; p24:=l24*r24; p25:=l25*r25; p26:=l26*r26; p27:=l27*r27; p28:=l28*r28; p29:=l29*r29; p30:=l30*r30; p31:=l31*r31; p32:=l32*r32; p33:=l33*r33; p34:=l34*r34; p35:=l35*r35; p36:=l36*r36; p37:=l37*r37; p38:=l38*r38; p39:=l39*r39; p40:=l40*r40; p41:=l41*r41; p42:=l42*r42; p43:=l43*r43; p44:=l44*r44; p45:=l45*r45; p46:=l46*r46; p47:=l47*r47;
\end{lstlisting}
\par
\begin{lstlisting}[style=slp,caption={\SLP for the output forms~\({\mat{P}}\)},label=lst:canP]
z25:=p21-p35; z40:=p43+p9; z22:=p12+p9; z33:=p31+p19; z21:=p11-p40; z43:=p47-z25; z52:=z25-z40; z36:=p37-z33; z49:=z33+z21; z46:=z52-z49; z27:=p23-z46; z44:=p8+z27; z50:=z49-z44; z30:=p27-p25; z34:=p34-z50; z41:=p45+z34; z24:=p16+z50; z32:=p3+z30; z45:=z52+z30; z29:=p26+z24; z42:=p46+z24; z31:=p29+z32; z23:=z32-p13; z10:=p18-z40-p17; z17:=z44-z22-p33; C[4,4]:=p15+p40+z22; z28:=p24-z45; z37:=p39+z45; z48:=z42-z10; z51:=z10-z23-z46; z35:=p36-z36; z26:=p22+z36; z38:=p4-z28; z19:=p28+z27+z48; z11:=p0-p25+z51; z18:=p14-z21+z48+z51; z39:=p42+z18; z20:=p2+z18+z22; z12:=z51-z17-p20; z13:=p1-z48+z37; z47:=z19-z20; z14:=z39-z11; C[3,1]:=z39+z41; C[2,3]:=z19+z29; z15:=z35-z29; C[1,2]:=z26-z12; C[1,1]:=z11-z31; z16:=z31-z26; C[4,2]:=p5+p17+z12+z13; C[2,2]:=z13-z35; C[3,4]:=p7-p35-z22+z41-z38; C[3,2]:=p10-p19-z41-z12+z13-z38; C[3,3]:=z20+z38; C[4,3]:=p32+z42+z28+z47; C[4,1]:=z23+z34+p44+z14; C[2,4]:=p30+z46-z37+z15; C[2,1]:=p38+z14+z15+z43; C[1,3]:=p6-z47-z43+z16; C[1,4]:=z17-z16-p41;
\end{lstlisting}

These straight-line programs require:
\begin{itemize}
\item~\(55\) additions/subtractions and two halvings for~\({\mat{L}}\);
\item~\(55\) additions/subtractions and two halvings for~\({\mat{R}}\);
\item~\(90\) additions/subtractions for~\({\mat{P}}\).
\end{itemize}
This gives~\(204\) linear operations, namely~\(200\)
additions/subtractions and~\(4\) halvings, in addition to the~\(48\)
bilinear products.
\par
For each construction below, let~\(F_{n}\) denote the total scalar
arithmetic cost at matrix order~\({n=4^{k}}\), including any input and
output conversions.
The full recursion satisfies~\({F_{1}=1}\)
and~\({F_{n}=48F_{\frac{n}{4}}+204{\left(\frac{n}{4}\right)}^2}\). This is the
recurrence of~\cref{lem:master-thm}
with~\({T(n)=F_{n}}\),~\({{a=48},{b=4},{c=204}}\) and~\({\beta=2}\),
giving
\begin{equation}
	F_{n}=\frac{59}{8}n^{\log_{4}\!{48}}-\frac{51}{8}n^2.
\end{equation}
\section{Alternative basis}
From the work started in~\cite{Karstadt:2017aa,Beniamini:2019aa}, one
can further reduce the complexity bounds using some change of
basis~\({\mat{X}=\mat{X}_{\textup{alt}}\cdot\mat{X}_{\textup{cob}}}\)
for the three matrices of the \textsc{lrp} representation.
For such a factorization, we call~\({\mat{X}_{\textup{cob}}}\) a
{\emph{change of basis}} matrix, or, more precisely,
an {\emph{encoder}} if~\(\mat{X}\) in~\({\{\mat{L},\mat{R}\}}\),
and a {\emph{decoder}} otherwise~(\({\mat{X}=\Transpose{\mat{P}}}\)).
We also call~\(\mat{X}_{\textup{alt}}\) the {\emph{residuum}}, or the
{\emph{residual matrix}}.
\par
For the \FMMA{4}{4}{4}{48} case, the change of basis
matrices~\(\mat{X}_{\textup{cob}}\) have~\(16\) columns and any number
of rows between~\(16\) and~\(47\).
This number of rows (and of columns of the
residuum~\(\mat{X}_{\textup{alt}}\)) is
denoted by the {\emph{common inner dimension, \({d}\)}, of the alternative
  basis}\footnote{Note that the change of basis matrices themselves
  could be further factorized; we do not consider this here as we
  focus on the optimization of the residuum~\(\mat{X}_{\textup{alt}}\).}.
\par
On the one hand, these change of bases incurs a theoretical negligible
complexity overhead ranging from~\(\bigO{n^2\log{n}}\) in the square
case, to~\({\bigO{n^{\log_{4}\!{47}}}}\) in the most rectangular case.
On the other hand, the resulting~\({\mat{X}_{\textup{alt}}}\) matrices can
then require fewer operations than the initial ones, and thus reduce
the dominant term of the complexity bound in~\({\bigO{n^{\log_{4}\!{48}}}}\).
The effect of the number of operations via these alternative basis is
the following: if the change of basis matrices have~\(d\) rows and the
alternative residual algorithms require now a total of~\(t\) operations,
the dominant term of the complexity bound is modified to be~\({\left(1+\frac{t}{48-d}\right)n^{\log_{4}\!{48}}}\).
\par
In the following we explore the square~\(\matrixsize{16}{16}\) and
most rectangular~\(\matrixsize{47}{16}\) cases,~\({d=16}\)
or~\({d=47}\), as well as intermediate encodings with inner
dimensions~\({d=24}\) or~\({d=32}\).
\subsection{Alternative basis with square change of basis (inner
  dimension d=16)}
Following~\cite{Karstadt:2017aa}, we present in this section
alternative bases with square change of basis, derived from the
matrices of~\cref{eq:444LRP,eq:canLRP}.
This is a factorization of each one of
the~\({48{\times}16}\)~\({\mat{L},\mat{R},\Transpose{\mat{P}}}\)
matrices into~\({48{\times}16}\) by~\({16{\times}16}\) matrix
products~\({\mat{X}=\mat{X}_{\textup{16alt}}\cdot\mat{X}_{\textup{16cob}}}\),
for~\(\mat{X}\) in~\({\{\mat{L},\mat{R},\Transpose{\mat{P}}\}}\).
\subsubsection{Sparse square alternative basis}
For the sparsest matrices~\cref{eq:444LRP}, we were able to find such decompositions with both~\(\mat{L}_{\textup{16alt}}\) and~\(\mat{R}_{\textup{16alt}}\) having~\(131\) non-zero elements (including~\(3\) non-units) and~\(\mat{P}_{\textup{16alt}}\) with~\(156\) non-zero elements (including~\(100\) non-units).
The associated change of basis matrices are such that:~\(\mat{L}_{\textup{16cob}}\) has~\(68\) non-zero elements (including~\(18\) non-units),~\(\mat{R}_{\textup{16cob}}\) has~\(65\) non-zero elements (including~\(13\) non-units) and~\(\mat{P}_{\textup{16cob}}\) has~\(70\) non-zero elements (including~\(19\) non-units).
\par
Finally we found straight-line programs for these matrices.
They require:
\(46\) additions for~\(\mat{L}_{\textup{16alt}}\);
\(46\) additions for~\(\mat{R}_{\textup{16alt}}\);
\(85\) additions and~\(8\) scalings (binary shifts) for~\(\mat{P}_{\textup{16alt}}\).
\(29\) additions and~\(2\) scalings (binary shifts) for~\(\mat{L}_{\textup{16cob}}\);
\(29\) additions and~\(2\) scalings (binary shift) for~\(\mat{R}_{\textup{16cob}}\);
\(33\) additions and~\(4\) scalings (binary shifts) for~\(\mat{P}_{\textup{16cob}}\).
\par
The number of operations for the alternative basis thus sums up to~\(185\) operations, and to~\(99\) operations for the square change of basis.
Overall, this gives a theoretical complexity bound of:
\begin{equation}
\left(1+\frac{185}{48-16}\right)n^{\log_4{\!48}}
+\left(\frac{99}{48-16}\right)n^2\log_2(n)
-\frac{185}{48-16}n^2
\approx 6.78125n^{2.7925}.
\end{equation}
\subsubsection{Canonical square alternative basis}
We factor the original canonical matrices of~\cref{eq:canLRP} as
\begin{equation}
\mat{L}=\mat{L}_{\textup{16alt}}\mat{L}_{\textup{16cob}},\quad
\mat{R}=\mat{R}_{\textup{16alt}}\mat{R}_{\textup{16cob}},\quad
\mat{P}=\mat{P}_{\textup{16cob}}\mat{P}_{\textup{16alt}}.
\end{equation}
Both~\(\mat{L}_{\textup{16alt}}\) and~\(\mat{R}_{\textup{16alt}}\) have~\(146\) non-zero elements, including~\(2\) non-units;~\(\mat{P}_{\textup{16alt}}\) has~\(250\) non-zeros, including~\(66\) non-units. The two input conversion matrices each have~\(78\) non-zeros, including~\(36\) non-units, and the output conversion has~\(60\) non-zeros, all units.
The corresponding straight-line programs require:
\begin{itemize}
\item~\(44\) additions/subtractions for each input residuum;
\item~\(83\) additions/subtractions, one halving and one doubling for the output residuum;
\item~\(36\) additions/subtractions and \(5\) doublings/halvings for each input conversion;
\item~\(27\) additions/subtractions for the output conversion.
\end{itemize}
Thus the residuum uses~\(173\) operations and the conversions
use~\(109\).
Let~\(K_{k}\) denote the arithmetic cost of the encoded residua
after~\(k\) recursion levels, including its scalar products but
excluding the global conversions.
It satisfies~\({K_0=1}\) and~\({K_k=48K_{k-1}+173\,16^{k-1}}\).
Writing~\({T(4^k)=K_k}\) gives the recurrence of~\cref{lem:master-thm} with~\({{a=48},{b=4},{c=173}}\) and~\({\beta=2}\).
The complete cost is
\begin{equation}
F_n=K_k+109k16^{k-1},\qquad n=4^k.
\end{equation}
Solving the residuum recurrence and adding this conversion cost gives
the exact count
\begin{equation}\label{eq:altsqrbasecomp}
F_n=\frac{205}{32}n^{\log_4 48}
+\frac{109}{16}n^2\log_2 n-\frac{173}{32}n^2.
\end{equation}
All six matching matrices and programs are available in the
\plinopt repository as
\plinoptdata{data/4x4x4_48_204-16{ALT,CoB}_{L,R,P}.s{ms,lp}}
(i.e.\ with prefix \nolinkurl{4x4x4_48_204}, residuum suffixes
\nolinkurl{-16ALT_{L,R,P}}, and conversion suffixes
\nolinkurl{-16CoB_{L,R,P}}).
In particular, the output conversion is composed on the left of the
output residuum.
We provide the conversion programs
in~\cref{lst:canCoBL,lst:canCoBR,lst:canICoB}, with~\(i_{j}\) the input
variables and~\(o_{j}\) the outputs. Exact replay, for example with
\plinopt's \texttt{SLPchecker}, reconstructs their matrices; the
factorization
identities~\({\mat{X}_{\textup{16alt}}=\mat{X}\cdot\mat{X}_{\textup{16cob}}^{-1}}\)
then determine the square-basis residua.
\begin{lstlisting}[style=slp,caption={\SLP
    for~\(\mat{L}_{\textup{16cob}}\)},label=lst:canCoBL]
t10:=i14+i13; t11:=i2+i14; t12:=(i0+i3)/2; t13:=(i9+i10-t10)/2; t14:=i7+i4; t15:=t12+t14/2; t16:=i6+i5; o5:=i1+t11-i14-i3; o0:=t16-i7; o14:=t13+i14; o10:=t14-i6; o7:=i6-t11; o3:=t15-i6; o9:=i1+t15; o6:=t15-i8; o1:=o0+i15-o5-t10; o4:=(i7-i4)/2+t10+t12+i12+o5-t16; o12:=t13-o5; o13:=o14+i13-o0; o15:=t15*2+i1+i10-t13-i11-i6-i8; o2:=i11; o8:=i6; o11:=i7;
\end{lstlisting}
\begin{lstlisting}[style=slp,caption={\SLP
    for~\(\mat{R}_{\textup{16cob}}\)},label=lst:canCoBR]
t10:=i7-i11; t11:=i10+i11; t12:=(i2-i14)/2; t13:=(i9-i5-t10)/2; t14:=i0-i12; t15:=t12+t14/2; t16:=i8-i4; o5:=i14-i11+t11-i6; o0:=i12+t16; t17:=t13-i9; o10:=t14-i8; o7:=t11-i8; o3:=t15-i8; o9:=t15-i6; o6:=i1+t15; o1:=o5+i15-t10-o0; o4:=o5+t12+i3-t16-t10-(i12+i0)/2; o12:=t13+o5; o14:=i11-t13; o15:=t17+i1+t15*2-i13-i8-i6; o13:=t17+i5-o0; o2:=i13; o8:=i8; o11:=i12;
\end{lstlisting}
\begin{lstlisting}[style=slp,caption={\SLP for~\(\mat{P}_{\textup{16cob}}\)},label=lst:canICoB]
o8:=i7-i12; z11:=i13+i11; z12:=i11-i5; o0:=i5-i4; o1:=i4-i11; o6:=i5+i4; o12:=i7-o0; z14:=i15-i7; o10:=z11-i10; z10:=i10-i11; o9:=z10-o8-i2; o11:=z10+o8; z13:=z11-o6; o3:=i0-z12; o15:=i8+i7+z12; o7:=i1-i5-z14; o5:=z14-i4; o14:=i6-z13; o13:=i3-o1-i7; o4:=i14-i15+o12-i9; o2:=z13+i9;
\end{lstlisting}
\subsection{An intermediate canonical encoding with inner dimension d=24}\label{ssec:r24}
An encoding with inner dimension~\({d=24}\) gives a different trade-off for the same original canonical matrices. Let the input residua~\({\mat{L}_{\textup{24alt}},\mat{R}_{\textup{24alt}}}\) have size~\({48\times24}\), the output residuum~\(\mat{P}_{\textup{24alt}}\) size~\({24\times48}\), the input encoders~\({\mat{L}_{\textup{24cob}},\mat{R}_{\textup{24cob}}}\) size~\({24\times16}\), and the output decoder~\(\mat{P}_{\textup{24cob}}\) size~\({16\times24}\). They satisfy
\begin{equation}
\mat{L}=\mat{L}_{\textup{24alt}}\mat{L}_{\textup{24cob}},\qquad
\mat{R}=\mat{R}_{\textup{24alt}}\mat{R}_{\textup{24cob}},\qquad
\mat{P}=\mat{P}_{\textup{24cob}}\mat{P}_{\textup{24alt}}.
\end{equation}
This is a redundant rectangular encoding, rather than an invertible
square change of basis. The residua require~\({32+32+58=122}\)
additions/subtractions and no scaling gates. Their non-zero counts
are~\({117,117,200}\), with~\({1,1,36}\) non-units. The encoders each
use~\(37\) additions/subtractions and~\(2\) halvings; the decoder
uses~\(70\) additions/subtractions and~\(3\) halvings, for~\(151\)
conversion operations in total.

The six \SLP and \textsc{sms} files are available
as \plinoptdata{data/4x4x4_48_204-24{ALT,CoB}_{L,R,P}.s{ms,lp}}
in the \plinopt repository,
using the same prefix and suffix conventions as the square case.
We give the conversion programs
in~\cref{lst:24CoBL,lst:24CoBR,lst:24ICoB}, with~\(i_{j}\) the input
variables and~\(o_{j}\) the outputs. Exact replay, for example with
\plinopt's \texttt{SLPchecker}, reconstructs their matrices; the
factorization
identities~\({\mat{X}_{\textup{24alt}}=\mat{X}\cdot\mat{X}_{\textup{24cob}}^{-1}}\)
then determine the square-basis residua.

\begin{lstlisting}[style=slp,caption={\SLP
    for~\(\mat{L}_{\textup{24cob}}\)},label=lst:24CoBL]
t10:=i0+i3; t11:=i4+i7; t18:=i9+i10; t21:=i6-i14; o20:=i1+t10; o6:=i6+i5-i7; o21:=i1+i13-i9; o19:=i10-t21; t12:=(t10+t11)/2; t13:=(t18+i13+i14)/2; t14:=i8-t12; t15:=t18-t13; o10:=t13-o6; o4:=i12+t12-t11; t16:=i11+t14; t17:=t15-i1; o5:=i15+o4; o2:=o19-t16; o14:=o21-t16; o11:=i2-t17; o15:=t21-t17; o18:=o10+t14+o4; o1:=i10-t16-t12; t19:=i3-o11; o8:=t15-o1; o9:=t15-t19; t20:=o4-t19; o16:=t20-t14; o17:=o10+t20; o22:=o18-t20; o23:=t16-o22; o0:=i6; o3:=i7; o7:=i11; o12:=i12; o13:=i8;
\end{lstlisting}
\begin{lstlisting}[style=slp,caption={\SLP
    for~\(\mat{R}_{\textup{24cob}}\)},label=lst:24CoBR]
t10:=i14-i2; t11:=i12-i0; t18:=i5-i9; t21:=i8-i11; o20:=i6+t10; o6:=i8+i12-i4; o21:=i6+i5+i7; o19:=i9+t21; t12:=(t10+t11)/2; t13:=(t18+i11-i7)/2; t14:=i1-t12; t15:=t18-t13; o10:=t13-o6; o4:=t11+i3-t12; t16:=t14-i13; t17:=i6+t15; o5:=o4-i15; o2:=t16-o19; o14:=o21-t16; o11:=i10-t17; o15:=t17-t21; o18:=t14-o4-o10; o1:=t12+t16-i9; t19:=i14+o11; o8:=t15-o1; o9:=t15+t19; t20:=o4+t19; o16:=t14+t20; o17:=o10+t20; o22:=o18+t20; o23:=o22-t16; o0:=i8; o3:=i12; o7:=i13; o12:=i3; o13:=i1;
\end{lstlisting}
\begin{lstlisting}[style=slp,caption={\SLP for~\(\mat{P}_{\textup{24cob}}\)},label=lst:24ICoB]
t10:=i23+i22; t21:=i20-i16; t27:=i23-i15; t32:=i19+t10; t11:=(i21-i18)/2; t12:=(t27+t21-i17)/2; t13:=(i14-t32)/2; t14:=i10-t32-t13; t15:=t27-t12; t16:=i23+t13; t17:=i12-t11; t18:=i11-t13-t17; t19:=i19+t11-t15; t20:=i7+t18; t22:=t21-t14+i9; t23:=i8+t11-t14-i17-i21; o15:=i18-i6; t24:=t20-t22; t25:=i0+t10-t15-t22; t26:=i5-t12-i14; t28:=t26-t24; o11:=i21-i1-t19-t19; t29:=t20-t25; t30:=t26-t25; o10:=t19+t28; o8:=t28-t19; t31:=t28+t25; t33:=i6+i23-t17-t12; t34:=t33+t19-t18; t35:=t10-t29-t33; o12:=t35-i13; o14:=t33+i20-t29; o0:=o8-i9-t35; o7:=i3+i23-t23-t34; t36:=o0-t30; o3:=t36+i22+t16-i2; o2:=t30-t36; o5:=t23-t31; o6:=t34+t31; o4:=t31-t34; o1:=t16-t30; o9:=t16+t23-i8-i5+t24; o13:=o5+i4-o1;
\end{lstlisting}

Taking tensor powers of the three factorization identities proves the
recursive construction, up to coordinate
permutations. With~\({n=4^k}\), let~\(K_k\) again count the encoded
residuum alone, including its scalar products. It satisfies
\begin{equation}
K_0=1,\qquad K_k=48K_{k-1}+122\,24^{k-1}.
\end{equation}
Writing~\({T(4^k)=K_k}\) gives the recurrence of~\cref{lem:master-thm} with~\({{a=48},{b=4},{c=122}}\) and~\({\beta=\log_{4}{24}}\). Adding the global tensor-product conversion cost gives
\begin{equation}
F_{n}=K_{k}+151\sum_{j=0}^{k-1}16^{k-1-j}24^{j},\qquad n=4^{k}.
\end{equation}
Solving the residuum recurrence and evaluating the geometric sum yields
\begin{equation}\label{eq:alt24comp}
F_{n}=\frac{73}{12}n^{\log_{4}\!{48}}
+\frac{331}{24}n^{\log_{4}\!{24}}-\frac{151}{8}n^{2}.
\end{equation}
This algorithm performs the conversions globally, rather than afresh inside each recursive call. At one level it costs~\({48+122+151=321}\) operations, compared with~\(252\) for the ordinary program and~\(304\) for the square encoding. The smaller asymptotic coefficient therefore does not imply an isolated~\({{4}\times{4}}\) cost improvement or a measured hardware speedup.
\subsection{An intermediate canonical encoding of P with inner dimension d=32}
We now look at the contributions, in terms of operations, of the
change of basis of each of~\({\mat{L}, \mat{R}}\) and~\(\mat{P}\) separately.
We have seen in~\cref{ssec:r24} that, with inner dimension~\({d=24}\),
their respective contributions to
the constant in the leading term are~\({\frac{32}{48-24}=1+\frac{1}{3}}\), \({\frac{32}{48-24}=1+\frac{1}{3}}\) and~\({\frac{58}{48-24}=2+\frac{5}{12}}\).
\par
We have found an encoding with inner dimension~\({d=32}\) that gives a lower contribution for the same original canonical output matrix~\(\mat{P}\).
An output residuum~\(\mat{P}_{\textup{32alt}}\) of
size~\({32\times{48}}\)
and an output decoder~\(\mat{P}_{\textup{32cob}}\)
of~\({16\times{32}}\), that
satisfy~\({\mat{P}=\mat{P}_{\textup{32cob}}\mat{P}_{\textup{32alt}}}\).
This residuum require~\({32}\) additions/subtractions and no scaling
gates; this is a contribution of only~\({\frac{32}{48-32}=2}\).
Their non-zero counts is~\({64}\), with no non-units.
The decoder uses~\(63\) additions/subtractions.
The two \SLP files are available in the \plinopt repository
as \plinoptdata{data/4x4x4_48_204-32{ALT,CoB}_P.slp}.
\par
We give the conversion program
in~\cref{lst:32ICoB}, with~\(i_{j}\) the input
variables and~\(o_{j}\) the outputs. Exact replay, for example with
\plinopt's \texttt{SLPchecker}, reconstructs the associated matrix; the
factorization identity then determine the square-basis residuum.
\begin{lstlisting}[style=slp,caption={\SLP for~\(\mat{P}_{\textup{32cob}}\)},label=lst:32ICoB]
w10:=i15+i18; w16:=i11-w10; w33:=i8+w10-i1; w34:=w33+w10-i16; w35:=w34-i30; w11:=i24+w34; w12:=i26+i16-w11; w13:=i18-w16-w11; o15:=i23+i19; w14:=i19-w35; w15:=i21-w33; w17:=w13-w11-w15; w18:=i1-i3+w14-w13; w19:=i7+i8+w13; w20:=w17+w33-i22; w21:=w35+w18; w22:=i25+w20; w23:=w17-i5-w19; w24:=i29-w15; o5:=i28-w21-w15; w25:=w19-w20; w26:=w19-i0; w27:=w25-i6; o11:=w14-i14; o8:=w14-w23; o10:=i4-w23; w28:=w19-w12; w29:=o10-i17; w30:=w26-o8; o6:=i17-w22; w31:=o5-w22; o1:=w24-w28; o0:=w26+w27; w32:=w24+w27; o14:=i9-w20-w29; o9:=i20-w18-w28; o13:=w28+i10-w21; o12:=i31-w25-w30; o7:=i2-w31; o4:=w16+i13-w31-w18-w30; o2:=w29-i12-w16-w32; o3:=w32+w12-i27;
\end{lstlisting}
Using the alternative bases of~\cref{ssec:r24} for~\(\mat{L}\) and~\(\mat{R}\), and the one of~\cref{lst:32ICoB} for~\(\mat{P}\) with~\({n=4^{k}}\),
let~\(K_{k}\) again count the encoded residuum alone, including its
scalar products. It satisfies
\begin{equation}
K_{0}=1,\qquad K_{k}=48K_{k-1}+(32+32)24^{k-1}+32\,32^{k-1}.
\end{equation}
With~\({T(4^{k})=K_{k}}\), the two forcing terms have the form of~\cref{lem:master-thm} with~\({{a=48},{b=4}}\) and respectively~\({(c,\beta)=(64,\log_{4}{24})}\) and~\({(32,\log_{4}{32})}\); their contributions add by linearity. Adding the global conversions for~\({\mat{L}_{\textup{24cob}},\mat{R}_{\textup{24cob}}}\) and~\(\mat{P}_{\textup{32cob}}\) gives
\begin{equation}
F_{n}=K_{k}+(39+39)\sum_{j=0}^{k-1}16^{k-1-j}24^j
	+63\sum_{j=0}^{k-1}16^{k-1-j}32^j,\qquad n=4^{k}.
\end{equation}
Solving the residuum recurrence and evaluating the geometric sums yields
\begin{equation}\label{eq:alt24_24_32comp}
F_{n}=\frac{17}{3}n^{\log_{4}\!{48}}
+\frac{31}{16}n^{\log_{4}\!{32}}
+\frac{85}{12}n^{\log_{4}\!{24}}
	-\frac{219}{16}n^{2}.
\end{equation}
\subsection{Sparse and canonical rectangular decompositions with inner
dimension d=47}
Following~\cite{Beniamini:2020aa}, we finally consider a rectangular
factorization with inner dimension~\({d=47}\). In the original canonical
matrix convention of~\cref{eq:canLRP}, using zero-based rank
indices, the rows and columns satisfy:
\begin{equation}
l_{44}=l_{5}+l_{18}+l_{34},\qquad
r_{30}=r_{7}-r_{21}+r_{39},\qquad
	c_{3}=-c_{13}+c_{29}.
\end{equation}
Let~\({\mat{L}_{h},\mat{R}_{h}}\) remove rows~\({44,30}\), respectively, and let~\(\mat{P}_{h}\) remove column~\(3\), retaining all other coordinates in ascending order. The factorizations are
\begin{equation}
\mat{L}=\mat{L}_{\textup{47alt}}\mat{L}_h,\quad
\mat{R}=\mat{R}_{\textup{47alt}}\mat{R}_h,\quad
\mat{P}=\mat{P}_h\mat{P}_{\textup{47alt}}.
\end{equation}
Here the input residua put the retained entries back at their original rank indices and reconstruct the omitted rows using respectively~\({\vec{e_5}+\vec{e_{18}}+\vec{e_{34}}}\) and~\({\vec{e_{7}}-\vec{e_{21}}+\vec{e_{38}}}\). The output residuum has a unit column at the retained position of each original column, except that original column~\(3\) is~\({-\vec{e_{12}}+\vec{e_{28}}}\). Thus all~\(48\) rank products keep their original pairing. These relations concern the displayed original matrices, not the sign-adjusted listings above.
\par
The algorithm defined by
these~\({\mat{L}_{\textup{47alt}},\mat{R}_{\textup{47alt}},\mat{P}_{\textup{47alt}}}\)
matrices, can be realized with straight-line programs with,
respectively,~\({2,2}\) and~\(2\) additions (for the
latter,~\(\Transpose{\mat{P}}_{\textup{47alt}}\) can obviously be
realized with~\(1\) additions, and
therefore~\(\mat{P}_{\textup{47alt}}\) requires~\({1+(48-47)=2}\)
additions, by the transposition principle).
\par
This gives a theoretical complexity bound of:
\({\left(1+\frac{2+2+2}{48-47}\right)n^{2+\log_4{\!{3}}}+o\!\left(n^{2+\log_4{\!3}}\right)\approx{7n^{2.7925}}}\),
worse than that of the square alternative basis case
in~\cref{eq:altsqrbasecomp,eq:alt24comp,eq:alt24_24_32comp}.
Even looking at encoders and decoders separately,
we see that the contribution of~\(\mat{L}_{\textup{47alt}}\)
and~\(\mat{L}_{\textup{47alt}}\) to the leading term constant is
twice~\(\frac{2}{1}\),
while those of~\({\mat{L}_{\textup{16alt}},\mat{R}_{\textup{16alt}},
\mat{L}_{\textup{24alt}},\mat{R}_{\textup{24alt}}}\)
are strictly below~\(2\) (namely~\({1+\frac{3}{8}}\) and~\({1+\frac{1}{3}}\)).
Note that the contribution of~\(\mat{P}_{\textup{47alt}}\) is also~\(
{2=\frac{2}{1}}\), equal to that of~\({\mat{P}_{\textup{32alt}}}\) at~\({2=\frac{32}{16}}\).
But the cost of the recursive application of the decoder~\(\mat{P}_{\textup{47cob}}\) is larger than that of~\(\mat{P}_{\textup{32cob}}\).
\begin{remark}
The analogous construction with inner dimension~\({d=47}\) for the
sparse variant matrices, also has leading coefficient~\(7\).
This comparison concerns these constructions with inner dimension~\({d=47}\), not all rectangular encodings.
\end{remark}
\section{Conclusion}
Combining geometric transformations and straight-line program
optimization gives a fast algorithm for multiplying
two~\(\matrixsize{4}{4}\) matrices using~\(48\) multiplications
and~\(204\) other operations (addition, subtraction or scaling by a
constant)
over any ring containing an inverse of~\(2\).
Applied recursively, this algorithm reaches a cost bound with leading
term constant~\({7+\frac{3}{8}}\).
Using alternative basis, this
constant can be reduced to~\({6+\frac{7}{16}}\) with square change of
basis,
and to~\({5+\frac{2}{3}}\) with rectangular ones, as summarized
in~\cref{tab:performance_comparison}.
All the presented variants use~\(48\) bilinear products; the encoded
bounds include the
lower-order conversion costs. These are attained arithmetic bounds,
without a claim of optimality or measured runtime improvement.
\par
However, a complete geometric understanding of this problem remains open.
Thus, further improvements for this problem could still be found
(e.g., more efficient \SLPs could exist, even for the \textsc{lrp}
representations given here, or within the family
of~\FMMA{4}{4}{4}{48} algorithm presented in~\cite{Li:2026aa}).
\par
Future work will also focus on numerical experiments to study the
efficiency of these algorithms and thus the practical benefits of the
theoretical optimizations presented here.
\bibliographystyle{plainurl}
\bibliography{mm444.bib}

@preamble{     "\def\noopsort#1{}"
}

@preamble{     "\newcommand{\issacproceedings}[2]{{ISSAC}'#1, Proceedings of the #1 International Symposium on Symbolic and Algebraic Computation, #2}"
}

@string{JSC = "Journal of Symbolic Computation"}

@string{journal:tcs = {Theoretical Computer Science}}

@string{publisher:acm = {Association for Computing Machinery}}

@string{journal:nm = {Numerische Mathematik}}

@string{journal:buams = {Bulletin of the American Mathematical Society}}

@string{publisher:springer = {Springer}}

@string{org:arxiv = {CoRR}}

@article{laderman:1976a,
	author = {Laderman, Julian David},
	cited-by = {pan:1984a, burichenko:2014},
	doi = {10.1090/S0002-9904-1976-13988-2},
	journal = journal:buams,
	month = jan,
	number = 1,
	pages = {126-128},
	title = {A noncommutative algorithm for multiplying~{${3}\times{3}$} matrices using~$23$ multiplications},
	volume = 82,
	year = 1976
}

@techreport{Karunaratne:2026aa,
	author = {Samurdhi Karunaratne and Anushka Idamekorala},
	institution = org:arxiv,
  eprint       = {2607.28676},
	number = {2607.28676},
	title = {55 Additions Suffice for 3x3 Matrix Multiplication at Rank 23},
	year = 2026,
	month = jul}

@techreport{Li:2026aa,
	author = {Xin Li and Yu Wang and Shenglong Hu},
	institution = org:arxiv,
	month = jul,
	eprint = {2607.15069},
	number = {2607.15069},
	title = {Substitution and quotient of the isotropy group action},
	year = 2026}

@article{groot:1978a,
	author = {Groote, de, Hans Friedrich},
	cited-by = {Blaser:2016aa, burichenko:2014},
	doi = {10.1016/0304-3975(78)90038-5},
	journal = journal:tcs,
	number = 2,
	pages = {1-24},
	title = {On varieties of optimal algorithms for the computation of bilinear mappings {I}. {T}he isotropy group of a bilinear mapping},
	volume = 7,
	year = 1978
}

@techreport{Beniamini:2020aa,
	author = {Beniamini, Gal and Cheng, Nathan and Holtz, Olga and Karstadt, Elaye and Schwartz, Oded},
	month = aug,
	institution = org:arxiv,
  eprint       = {2008.03759},
	number = {2008.03759},
	title = {Sparsifying the Operators of Fast Matrix Multiplication Algorithms},
	year = {2020}}

@inproceedings{Beniamini:2019aa,
	address = {Phoenix, Arizona, USA},
	author = {Beniamini, Gal and Schwartz, Oded},
	booktitle = {SPAA'19: Proceedings of the 31st annual ACM {S}ymposium on {P}arallel {A}lgorithms and {A}rchitectures},
	doi = {10.1145/3323165.3323188},
	editor = {Berenbrink, Petra and Scheideler, Christian},
	location = {Phoenix, Arizona, USA},
	month = jun # {~22-24},
	pages = {11-22},
	publisher = publisher:acm,
	title = {Fast Matrix multiplication via sparse decomposition},
	year = 2019
}

@techreport{alphaevolve,
	author = {Novikov, Alexander and V{\~u}, Ng{\^a}n and Eisenberger, Marvin and Dupont, Emilien and Huang, Po-Sen and Wagner, Adam Zsolt and Shirobokov, Sergey and Kozlovskii, Borislav and Ruiz, Francisco J. R. and Mehrabian, Abbas and Kumar, M. Pawan and See, Abigail and Chaudhuri, Swarat and Holland, George and Davies, Alex and Nowozin, Sebastian and Kohli, Pushmeet and Balog, Matej},
	institution = org:arxiv,
	eprint = {2506.13131},
	number = {2506.13131},
	title = {Alpha{E}volve: A coding agent for scientific and algorithmic discovery},
	year = 2025,
	month = may,
}

@article{DPS25:444,
  title={A non-commutative algorithm for multiplying 4x4 matrices using 48 non-complex multiplications},
  author={Dumas, Jean-Guillaume and Pernet, Cl{\'e}ment and Sedoglavic, Alexandre},
  year={2027},
  eprint       = {2506.13242},
  volume = 3,
  number = 1,
  month = mar,
  journal = "Journal of Experimental Mathematics",
  note = "To appear",
}

@techreport{Dumas:2026ac,
	author = {Dumas, Jean-Guillaume and Pernet, Cl{\'e}ment and Sedoglavic, Alexandre},
	eprint = {2603.18699},
	number = {2603.18699},
	institution = org:arxiv,
	month = mar,
	primaryclass = {cs.DS},
	title = {A more accurate rational non-commutative algorithm for multiplying 4x4 matrices using 48 multiplications},
	year = 2026}

@article{Morgenstern:1975:Linear,
    title = {The {Linear} {Complexity} of {Computation}},
    volume = {22},
    issn = {0004-5411, 1557-735X},
    doi = {10.1145/321879.321881},
    language = {en},
    number = {2},
    urldate = {2025-01-13},
    journal = {Journal of the ACM},
    author = {Morgenstern, Jacques},
    month = apr,
    year = {1975},
    pages = {184--194},
}

@misc{jgd:2024:plinopt,
	author = {Dumas, Jean-Guillaume and Grenet, Bruno and Pernet, Cl{\'e}ment and Sedoglavic, Alexandre},
	month = jan,
	note = {8.09~kSLOC},
	organization = {v2.3, 937cdb3},
	title = {{PLinOpt}, a collection of {C++} routines handling linear \& bilinear programs},
	url = {{https://github.com/jgdumas/plinopt}},
hal = "hal-04927443",
	year = 2024,
}

@article{Dumas:2026:autoaccurate,
  title = {Towards automated generation of fast and accurate algorithms for recursive matrix multiplication},
  author = "Jean-Guillaume Dumas and Cl\'ement Pernet and Alexandre Sedoglavic",
journal = JSC,
year = {2026},
month = "May--June",
volume = {134},
number = {102524},
issn = {0747-7171},
  hal = {hal-04995684},
eprint = {2506.19405},
  doi = {10.1016/j.jsc.2025.102524},
}

@inproceedings{Dumas:2026:addsf,
  crossref =      {2026:ISSAC:Vu},
  author = "Jean-Guillaume Dumas and Stefano Lia and John Sheekey",
  title = {Computational Explorations on the Tensor rank and the Additive complexity of Semifields},
pages = 19,
  doi = "10.1145/3815436.3815471",
  eprint = {2602.09577},
  hal = {hal-05501053},
}

@inproceedings{Karstadt:2017aa,
	acmid = {3087579},
	address = {New York, NY, USA},
	author = {Karstadt, Elaye and Schwartz, Oded},
	booktitle = {SPAA'17: Proceedings of the 29th ACM Symposium on Parallelism in Algorithms and Architectures},
	doi = {10.1145/3087556.3087579},
	editor = {Scheideler, Christian and Hajiaghayi, Mohammad},
	isbn = {978-1-4503-4593-4},
	location = {Washington, DC, USA},
	month = jul # {~24--26},
	pages = {101--110},
	publisher = {ACM Press},
	title = {Matrix Multiplication, a Little Faster},
	url2 = {https://github.com/elayeek},
	year = 2017,
}

@incollection{Boyar:2008:shortestSLP,
	address = {Berlin, Heidelberg},
	title = {On the {Shortest} {Linear} {Straight}-{Line} {Program} for {Computing} {Linear} {Forms}},
	volume = {5162},
	isbn = {978-3-540-85237-7 978-3-540-85238-4},
	issn = {0302-9743, 1611-3349},
	doi = {10.1007/978-3-540-85238-4_13},
	language = {en},
	urldate = {2026-06-18},
	booktitle = {Mathematical {Foundations} of {Computer} {Science} 2008},
	publisher = {Springer Berlin Heidelberg},
	author = {Boyar, Joan and Matthews, Philip and Peralta, René},
	editor = {Ochmański, Edward and Tyszkiewicz, Jerzy},
	year = {2008},
	note = {Series Title: Lecture Notes in Computer Science},
	pages = {168--179},
}

@techreport{Perminov:2025:parallelCSE,
	title = {Parallel {Heuristic} {Exploration} for {Additive} {Complexity} {Reduction} in {Fast} {Matrix} {Multiplication}},
	institution = org:arxiv,
	number = {2512.13365},
	eprint = {2512.13365},
	urldate = {2025-12-17},
	author = {Perminov, Andrew Igorevich},
	month = dec,
	year = {2025},
}

@article{Fawzi:2022aa,
    author = {Fawzi, Alhussein and Balog, Matej and Huang, Aja and Hubert, Thomas and Romera-Paredes, Bernardino and Barekatain, Mohammadamin and Novikov, Alexander and R. Ruiz, Francisco J. and Schrittwieser, Julian and Swirszcz, Grzegorz and Silver, David and Hassabis, Demis and Kohli, Pushmeet},
    doi = {10.1038/s41586-022-05172-4},
    id = {Fawzi2022},
    isbn = {1476-4687},
    journal = {Nature},
    month = oct,
    number = {7930},
    pages = {47--53},
    title = {Discovering faster matrix multiplication algorithms with reinforcement learning},
    url = {https://doi.org/10.1038/s41586-022-05172-4},
    volume = {610},
    year = {2022},
}

@article{Winograd:1977,
	author = {Winograd, Shmuel},
	journal = {La Recherche},
	month = nov,
	volume = 83,
	pages = {956--963},
	title = {La complexit{\'e} des calculs num{\'e}riques},
	year = 1977}

@article{strassen:1969,
    author = {Strassen, Volker},
    cited-by = {brent:1970, winograd:1971, pan:1984a, burichenko:2014},
    doi = {10.1007/BF02165411},
    journal = journal:nm,
    month = aug,
    number = 4, 
    pages = {354--356},
    publisher = publisher:springer,
    title = {{G}aussian elimination is not optimal},
    volume = 13,
    year = 1969,
}

@proceedings{2026:ISSAC:Vu,
  title =         {\issacproceedings{2026}{{Oldenburg}, {Germany}}},
  booktitle =     {\issacproceedings{2026}{{Oldenburg}, {Germany}}},
  logo      = "{ISSAC}'2026",
  editor =        {Thi-Xuan Vu},
  month =         jul,
  day = "13--17",
  location  = {Oldenburg, {Germany}},
  publisher =     {ACM Press}, address =       {New York},
  year =          {2026},
}

@techreport{Martensson:2026ab,
	author = {Erik M{\aa}rtensson and Paul Stankovski Wagner},
	institution = {Cryptology {ePrint} Archive},
	month = may,
	title = {On Why and How to Minimize the Arithmetic Complexity of Fast Matrix Multiplication Algorithms},
	url = {https://eprint.iacr.org/2026/849},
	year = 2026}
\end{document}